\documentclass{pasj00}

\begin{document}
\SetRunningHead{Matsushita et al.}{Metal distribution in the ICM of the Fornax cluster}
\Received{2000/12/31}
\Accepted{2001/01/01}

\title{Suzaku observation of the metallicity distribution in the intracluster
medium of the Fornax cluster}

\author{Kyoko \textsc{Matsushita}\altaffilmark{1}, Yasushi \textsc{Fukazawa}\altaffilmark{2}, John P.\
 \textsc{Hughes}\altaffilmark{3}, Takao \textsc{Kitaguchi}\altaffilmark{4}, \\Kazuo \textsc{Makishima}\altaffilmark{4},
 Kazuhiro \textsc{Nakazawa}\altaffilmark{5}, Takaya \textsc{Ohashi}\altaffilmark{6}, Naomi \textsc{Ota}\altaffilmark{7}\\,
Takayuki \textsc{Tamura}\altaffilmark{5}, Miyako \textsc{Tozuka}\altaffilmark{2}, Takeshi Go
\textsc{Tsuru}\altaffilmark{8}, Yuji \textsc{Urata}\altaffilmark{7}, \\and Noriko, Y.
\textsc{Yamasaki}\altaffilmark{5}}
  \altaffiltext{1}{Department of Physics, Tokyo University of Science,\\
 Kagurazaka, Shinjyuku-ku, Tokyo, 162-8601, Japan}
\email{matusita@rs.kagu.tus.ac.jp}
 \altaffiltext{2}{Department of Physical Science, 
 Hiroshima University, 1-3-1 Kagamiyama, \\Higashi-Hiroshima, Hiroshima
 739-8526, Japan}
\altaffiltext{3}{Department of Physics and Astronomy, Rutgers
 University, Piscataway, NJ 08854-8019, USA}
\altaffiltext{4}{Department of Physics, The
 University of Tokyo, 7-3-1 Hongo, Bunkyo-ku, Tokyo 113-0033, Japan}
\altaffiltext{5}{Institute of Space and Astronautical Science, Japan
 Aerospace Exploration Agency,\\ 2-1-1 Yoshinodai, Sagamihara, Kanagawa
 229-8510, Japan}
\altaffiltext{6}{Department of Physics, Tokyo Metropolitan University,
 Hachioji, Tokyo 192-0397, Japan}
\altaffiltext{7}{Institute of Physical and Chemical Research (RIKEN),
 2-1 Hirosawa, Wako, Saitama 351-0198, Japan}
\altaffiltext{8}{Department of Physics, Kyoto University, Sakyo-ku, Kyoto 606-8502,Japan}
\maketitle
\begin{abstract}
The metallicity distribution in the Fornax cluster was studied with the
XIS instrument onboard the Suzaku satellite. K-shell lines of O and Mg were resolved clearly,
and the abundances of O, Mg, Si, S and Fe were measured with good
accuracy.  The region within a  $4'$ radius of NGC 1399 shows approximately
solar abundances of Fe, Si and S, while the O/Fe and Mg/Fe abundance ratios
are about 0.4--0.5 and 0.7 in solar units.  In the outer region spanning radii
between $6'$ and $23'$, the Fe and Si abundances drop to 0.4--0.5 solar and show no
significant gradient within this region. The abundance ratios, O/Fe and
Mg/Fe, are consistent with those in the central region.  We also
measured the Fe abundance around NGC 1404 to be approximately solar, and the O,
Ne and Mg abundances to be 0.5--0.7 times the Fe level.  The significant 
relative enhancement of Fe within 130 kpc of NGC 1399 and in NGC 1404 
indicates  an origin in SN Ia, in contrast to the species O, Ne, and Mg which
reflect the stellar metallicity.  
The mass-to-light ratios for O and Fe within 130 kpc of NGC 1399
 are over an order of magnitude lower than those in
rich clusters, reflecting the metal enrichment history of this poor cluster.

\end{abstract}

\section{Introduction}

The metal abundances in the intracluster medium (ICM) and
the hot X-ray emitting interstellar medium (ISM) of early-type galaxies  provide important clues to
understand the metal enrichment history and evolution of 
galaxies.
The ICM contains a large amount of metals, which
are mainly synthesized by supernova (SN) in early-type galaxies (e.g., Arnaud et al.\
1992; Renzini et al.\ 1993). 

The ASCA satellite (Tanaka et al.\ 1994) first enabled us to measure the
distribution of Fe in the ICM\@ (e.g., Fukazawa et al. 2000; Finoguenov
et al. 2000; 2001).  The derived iron-mass-to-light ratio
(IMLR) is nearly constant in rich clusters and decreases toward poorer
systems (Makishima et al.\ 2001).  In individual clusters, the IMLR is
lower around the center (Makishima et al.\ 2001).  The Si abundance of the 
ICM was also measured by ASCA (Fukazawa et al.\ 1998;2000;
Finoguenov et al.\ 2000; 2001), however, since Fe and Si are both
synthesized in SN Ia and SN II, abundance results for these species
cannot clearly constrain the nucleosynthesis contribution from
different SN types.  In contrast, O and Mg are predominantly
synthesized in SNe II. Abundance measurements
spanning the range of species from O to Fe are therefore needed to
obtain unambiguous information on the formation history of 
massive stars.

XMM-Newton provided the means to measure O and Mg abundances in some systems,
but reliable results have been
obtained only for the central regions of very bright clusters or
groups of galaxies dominated by cD galaxies (e.g., Xu et al.\
2002, Finoguenov et al.\ 2002, Tamura et al.\ 2003, Matsushita et al.\
2003, 2006; Buote et al (2003)).  In general, the measured abundances of Si and Fe are 
similar, while the O abundance  is only about half  of the Fe abundance.
This indicates
that cD galaxies are an important source of Si and Fe which are mainly
synthesized in SN Ia.
However, our knowledge about the distributions of O and Mg in the ICM is still
poor, in contrast to the detailed measurement of the Fe and Si
abundances with XMM-Newton (e.g., Tamura et al.\ 2004). 

Regarding early-type galaxies, the metal abundances in the ISM give us
important information about the present metal supply into the ICM
through SN Ia and stellar mass loss.  In addition, O and Mg abundances
should reflect the stellar metallicity and enable us to directly look
into the formation history of these galaxies.

In this paper, based on Suzaku (Mitsuda et al.\ 2006) observations,
the abundances of O, Mg, Si, S and Fe of the ICM in the Fornax
cluster and those of O, Ne, Mg and Fe of the ISM in NGC 1404 are
discussed.  
The XIS (Koyama et al.\ 2006) instrument onboard Suzaku has good energy resolution at the O line energy
with low background, providing better sensitivity than the EMOS
onboard XMM-Newton. The EMOS also has a problem in measuring the Mg
abundance in somewhat fainter systems, due to a strong instrumental Al
line.  The Mg lines are particularly useful in cluster outskirts,
since the strong Galactic O line causes difficulty in the measurement
of O emission from nearby clusters.

The Fornax cluster is a nearby poor cluster with an ICM temperature
of 1.3-1.5 keV (e.g., Scharf et al.\ 2005).  
 The X-ray emission shows
asymmetric spatial distribution, and the cD galaxy, NGC 1399, is
offset from the center (Paolillo et al.\ 2002; Sharf et al.\ 2005), 
 which may be related to large scale dynamical
       evolution such as infall motions of galaxies into the cluster
       (Dunn \& Jerjen 2006).  Chandra observations suggests that 
 there may be relative motion between NGC 1399 and the ICM, and
the 2nd brightest elliptical galaxy, 
NGC 1404, is moving supersonically
 in the ICM (Scharf et al.\ 2005; Machacek et al.\ 2005).
For the ICM abundances, the Fe and Si abundances within $\sim 50$ kpc
of NGC 1399 were measured with XMM-Newton (Buote et al.\ 2002).
 However, due to the high background of XMM-Newton,
 the O and Mg abundances of the ICM were not determined.

ASCA detected excess hard X-ray emission from several groups of
galaxies including the Fornax cluster (Fukazawa et al.\ 2001; Nakazawa
2001). Dynamical motions in the cluster, driven by merger events,
          that generate a population of relativistic particles may be the
           origin of this emission component. Suzaku results on hard
           emission from Fornax will be presented in a separate article
           after a detailed study of detector background.

We adopt   solar abundances by Feldman (1992),
where the solar O and Fe abundances relative to H are $8.51\times
10^{-4}$ and  $3.24 \times10^{-5}$ by number. 
Recently, the solar photospheric abundances of C, N, O, and Ne decreased
by 0.2 dex, considering three-dimensional hydrostatic model atmospheres
and non-local thermodynamic equilibrium (Asplund 2005 and references
therein).
The new solar O and Fe abundances relative to H are 
4.90$\times 10^{-4}$ and $2.95\times 10^{-5}$, respectively (Lodders 2003).
The effect of the new solar abundances will be discussed  in Section 4.

 We adopt 19 Mpc (for H$_0$ = 72 km/s/Mpc) for the distance to the
Fornax cluster.  The observation of Cepheids with  Hubble Space
Telescope yielded the distance of 18.6$\pm 0.6$ Mpc (Madore et al. 1999).
Unless otherwise specified, errors are quoted at 90\% confidence

\section{Observations}

\begin{figure}
  \begin{center}
    \FigureFile(60mm,40mm){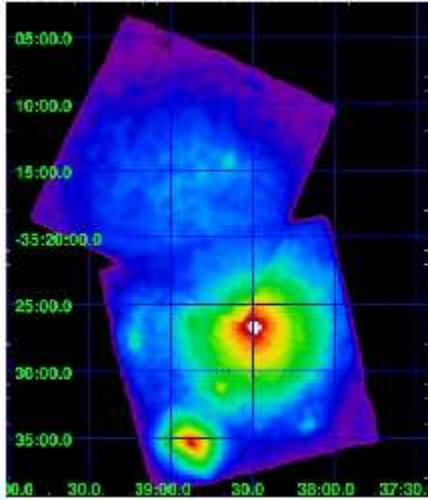}
  \end{center} \caption{The 0.3-5.0 keV Suzaku XIS image of the Fornax
  cluster. 
Data from the BI and FI detectors were combined.
 The difference of exposure time was corrected. 
}\label{fig:image}
\end{figure}

The Fornax cluster was observed twice as a part of the initial
performance verification of Suzaku.  The first observation (hereafter,
the central field) was carried out on September 2005 with
pointing direction $2'$ south and $1'$ east of NGC 1399.  The
second one (hereafter, the north field) was centered $13'$ north
and $4'$ east of NGC 1399, and was carried out on 2006 January.  Figure
\ref{fig:image} shows a 0.3-5.0 keV image for the two fields.
 The two
peaks in the central field correspond to the cD galaxy, NGC 1399, and
the elliptical galaxy, NGC 1404.  

Both the X-ray Imaging Spectrometers (XIS; Koyama et al.\ 2006) and
the Hard X-ray Detector (HXD) were operated in their nominal
modes.  We discarded the data taken with a cut-off rigidity less than
6 GV c$^{-1}$, or elevation angle less than $10^\circ$ from the earth
rim.  This yielded exposure times of 68 ks and 85 ks for the central
and the north fields, respectively. 
 We used the response matrix files
ae\_xi[0,2,3]\_segc\_rmf\_20060213.fit for the front side
illuminated (FI) detectors, XIS0, XIS2, and XIS3, and
ae\_xi1\_20060213c.rmf for the back side illuminated (BI) detector,
XIS1. 
The energy range analyzed here covers 0.3 to 5.0 keV for
the BI detector and 0.4 to 5.0 keV for the FI detectors.
 The on-axis auxiliary response files 
ae\_xi[0,1,2,3]\_xisnom4\_20060415.arf are used for the respective
detectors,   
since  below 5 keV, the energy dependence of the auxiliary
response file is smaller than the statistical errors.
We also ignored the energy
range between 1.82 and 1.84 keV in the spectral fit, since the
response matrix around the Si edge has some problem.

\section{Analysis and Results}

\subsection{Spectra of NGC 1399 and ICM}

\begin{figure}
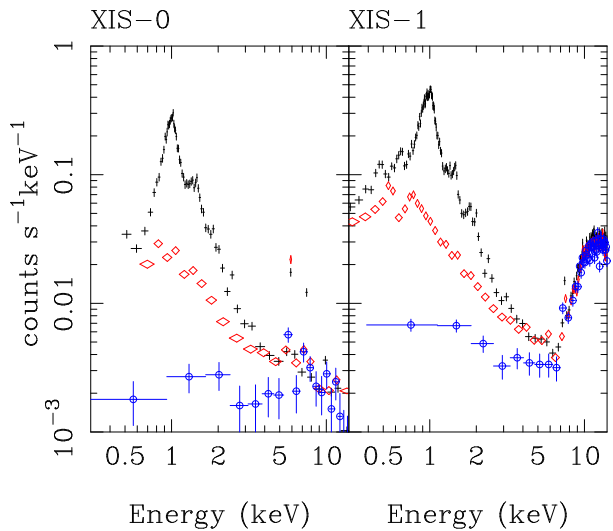

  \begin{center}
    \FigureFile(80mm,40mm){PASJ2938_figure2.ps} 
\end{center} 
    \caption{The raw spectra (black crosses) from XIS0 (FI) and XIS1 (BI)
	extracted from the outermost region, $r>16'$, in the north
	field. Spectra for the Lockman Hole (red diamonds) and the night
	earth (blue open circles) accumulated over the same detector area
	are shown for comparison. }\label{fig:raw_spec}
\end{figure}

The data from each XIS detector were accumulated within concentric
rings, centered on NGC 1399.  The spectrum of the bright galaxy NGC
1404 was separately analyzed, and a region with a radius of 4'
centered on this galaxy was masked out for this portion of the analysis.  As for the background,
we have to consider the non X-Ray background (NXB), the Cosmic X-Ray
Background (CXB), and the Galactic emission,  which arises from the 
 local hot bubble (LHB) and Milky Way halo (MWH).  Figure
\ref{fig:raw_spec} shows the spectra for the outermost ($r>16'$)
region of the Fornax cluster, compared with the blank-sky (Lockman
Hole observed on 2005 November) and the night-earth spectra
accumulated over the same region of the detector. In the low energy band the
Lockman-Hole data is not appropriate for the background, since the
temperature and intensity of the Galactic emission differ significantly with
sky region and the low-energy efficiency of the XIS instrument
has been declining since launch due to contamination on the XIS filters 
(Koyama et al.\ 2006).
Therefore, we subtracted the night-earth data as the non X-Ray
background, and the CXB and the Galactic emission are included as
models in the spectral fit.

We fitted the background-subtracted spectra with a model consisting of
4 components: a single-temperature vAPEC model (Smith et al.\ 2001) for the ICM, a
power-law model for discrete sources and the possible hard component,
a power-law model for the CXB, and an APEC model  for the LHB.
  (We exclude MWH component, since  as summarized in table 1 fits to the
             outermost region yield a zero normalization for it.)
 The temperature of the LHB was  fixed to 0.08 keV.
The spectra of the FI detectors (XIS0, XIS2, XIS3) were
fitted simultaneously.  The model spectra, except for the Galactic
emission, were subject to a common interstellar absorption $N_{\rm H}$,
fixed at the Galactic value, $1.3\times 10^{20}\rm{cm^{-2}}$.
  The ICM abundances of C and N were fixed
to solar, while those of the other species
                 were allowed to vary.  For the XIS1
(BI) spectra of the north field, the temperature and 
normalization of the Galactic emission were left free with 
abundance fixed to solar. For the other FI and BI spectra,
the parameters of the Galactic emission were fixed to the
best-fit values obtained with the BI detector for the $r>16'$ region in
the north field,   which   are summarized in table \ref{table:bgd}.
Within a radius of $4'$, we also applied a
two-temperature model for the ICM, where the metal abundances of the
two components were assumed to have the same value.

\begin{table}[t]
\caption{Parameters of background obtained with  the BI detector for  $r>16'$ in the north field}
\label{table:bgd}
 \begin{tabular}[t]{lll}
  LHB $kT$  (keV) & 0.08 & fixed \\
  MWH $kT$ (keV) & 0.20 & fixed \\
  ratio of normalization of MWH/LHB & 0.0& ($<0.06$) \\
  CXB Photon index & 1.4 & fixed\\
 \end{tabular}
\end{table}

As mentioned above, contamination has been building up 
on the XIS camera filters reducing the low energy efficiency. A
spectral model describing this effect has been developed.  We
assume the chemical composition of the contamination to have a
O/C number ratio of 1/6, and in the fits we allow the thickness of
the molecular layer to vary independently for each XIS detector.
Later we assess systematic uncertainties arising from our models
of the Galactic emission and contamination.

\begin{figure}
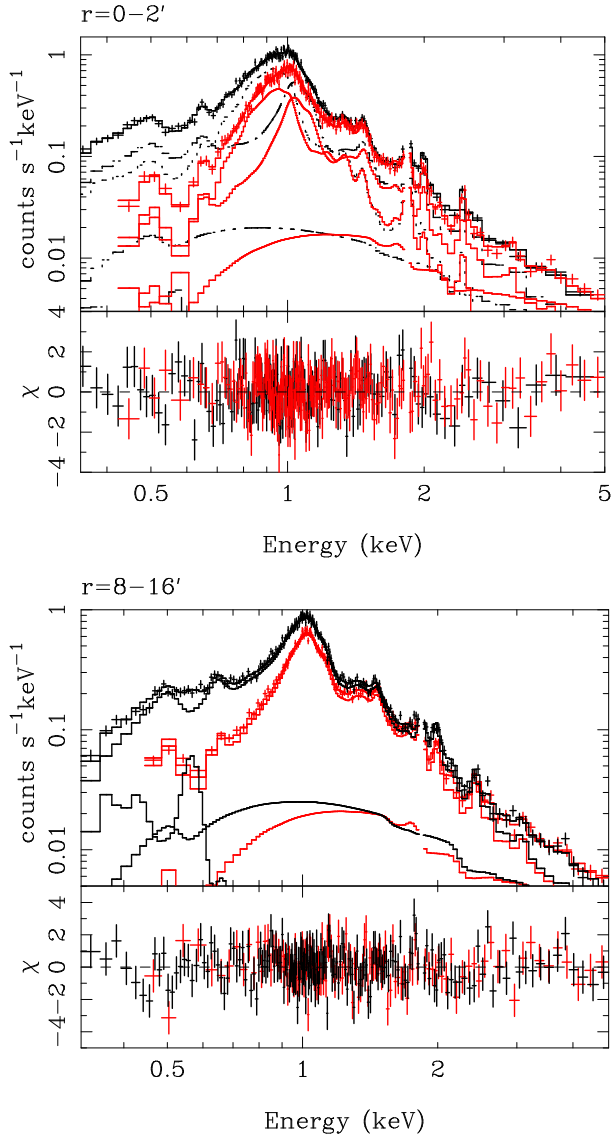

  \begin{center}
    \FigureFile(80mm,40mm){PASJ2938_figure3a.ps}
    \FigureFile(80mm,40mm){PASJ2938_figure3b.ps}
  \end{center}

  \caption{(left panel) The XIS0 (red) and XIS1 (black) spectra from
  $r<2'$, with the night earth data subtracted as the background,
  fitted with a 4-component model: two vAPEC models for the ICM, a
  power law for the CXB, and an APEC model for the Galactic emission.
  (right panel) The spectra at $8'<r<16'$ in the north
  field, fitted with the same model but with a single-temperature
  vAPEC model for the ICM.  Residuals are shown at the bottom of each
  panel.}\label{fig:spec}
\end{figure}

Table 2 summarizes the results of the spectral fits.  The BI and FI
detectors (in table 2 labeled XIS 1 and XIS 023, respectively)  have
given
generally consistent best-fit values for
temperature and abundance.  Within $r
= 4'$, the two-temperature model significantly improved the fit.
Figure \ref{fig:spec} shows representative spectra for the XIS0 and
XIS1 detectors.  
 Since K-shell lines of Ne are completely hidden in the Fe-L region,
we do not present the Ne abundance.
The Fe-L bump, H-like Mg, Si, and S lines are all
fitted well with this model.  The O lines are well modeled for the
data outside of $ r = 2'$.  However, for the inner region ($r <2'$),
the fit around the H-like O line of the XIS1 is not satisfactory: the $\chi^2$
value for the limited energy range 0.57--0.73 keV is 20 for 12 bins.
Adding another temperature component does not improve the fit in the
inner region.
For the FI detectors, as described in \S3.2 below,  the O abundance
may also have a systematic uncertainty at $r<2'$, since
the derived value of the column density of the contaminant is 
significantly  higher than the empirical model.

\begin{table*}[t]
\caption{Spectral fit results for NGC 1399 and the ICM.}\label{n1404}
\begin{center}
  \begin{tabular}[t]{llllllllllr}
r$^a$ &model & XIS  & kT &  EM ratio$^b$ & O & Mg & Si & S & Fe &$\chi^2$/dof \\
   &   &           & (keV) & &(solar) & (solar) & (solar) & (solar) &  (solar) &\\\hline
\multicolumn{10}{c}{The central field}\\\hline
0-2$'$ & 1T$^c$ & 1   &   0.99  && 0.18 &0.31 &0.49 &0.64 &0.51 &461/147\\
0-2$'$ & 2T$^d$ & 1   &  1.48$\pm$0.10&0.60$\pm$0.07&0.34$\pm$0.05 &0.97$\pm$0.13
   &1.14$\pm$0.13 &1.09$\pm$0.19 &1.18$\pm$0.16 &145/145\\
       &        &     & 0.82$\pm$0.01  \\
0-2$'$ & 1T & 023 &  1.00 && 0.24 &0.35 &0.54 &0.63 &0.60
   &848/256\\
0-2$'$ & 2T & 023 &  1.35$\pm$0.02&0.70$\pm$0.12&0.47$\pm$0.10 &0.64$\pm$0.07 &0.94$\pm$0.07 &0.92$\pm$0.11 &0.98$\pm$0.04 &270/253\\
       &   &    & 0.81$\pm0.02$ \\
2-4$'$ & 1T & 1 &  1.24$\pm$0.02 &&0.28$\pm$0.06 &0.38$\pm$0.11
   &0.68$\pm$0.09 &0.57$\pm$0.12 &0.62$\pm$0.03 &253/154\\
2-4$'$ & 2T & 1 &  1.41$\pm$0.05 &0.20$\pm$0.02&0.44$\pm$0.09 &0.88$\pm$0.17 &1.07$\pm$0.14 &0.83$\pm$0.18 &0.97$\pm$0.10 &164/152\\
       &  &  &  0.84$\pm$0.02\\
2-4$'$ & 1T & 023 & 1.27$\pm$0.01 &&0.30$\pm$0.08 &0.40$\pm$0.07
   &0.67$\pm$0.06 &0.76$\pm$0.08 &0.65$\pm$0.03 &356/258 \\
2-4$'$ & 2T & 023 &  1.47$\pm$0.11&0.23$\pm$0.08&0.45$\pm$0.11 &0.65$\pm$0.08 &0.90$\pm$0.09 &0.95$\pm$0.08 &0.90$\pm$0.10 &285/256\\
       &  &  &  0.96$\pm$0.10 \\
4-6$'$ & 1T & 1 &  1.33$\pm$0.02 &&0.17$\pm$0.06 &0.35$\pm$0.14
   &0.54$\pm$0.09 &0.59$\pm$0.14 &0.55$\pm$0.05 &157/123\\
4-6$'$ & 1T & 023 & 1.34$\pm$0.01 &&0.24$\pm$0.09 &0.49$\pm$0.08
   &0.67$\pm$0.07 &0.63$\pm$0.10 &0.56$\pm$0.03 &223/218\\
6-11$'$ & 1T & 1 &1.23$\pm$0.02 &&0.31$\pm$0.08 &0.40$\pm$0.10
   &0.32$\pm$0.09 &0.30$\pm$0.17 &0.42$\pm$0.04 &156/129\\
6-11$'$ & 1T & 023 &  1.31$\pm$0.02 &&0.20$\pm$0.08 &0.35$\pm$0.08 &0.37$\pm$0.08 &0.49$\pm$0.10 &0.40$\pm$0.03 &274/230\\
\hline
\multicolumn{10}{c}{The north field}\\\hline
8-16$'$ & 1T & 1 & 1.24$\pm$0.02 &&0.23$\pm$0.06 &0.36$\pm$0.10 &0.40$\pm$0.07 &0.37$\pm$0.11 &0.49$\pm$0.05 &200/174\\
8-16$'$ & 1T & 023 & 1.25$\pm$0.01 &&0.31$\pm$0.07 &0.36$\pm$0.04 &0.42$\pm$0.04 &0.49$\pm$0.06 &0.54$\pm$0.03 &789/651\\
16-23$'$ & 1T & 1 &  1.06$\pm$0.01 &&0.23$\pm$0.08 &0.33$\pm$0.14
   &0.44$\pm$0.10 &0.43$\pm$0.21 &0.43$\pm$0.04 &110/83\\
16-23$'$ & 1T & 023  & 1.06$\pm$0.01& &0.24$\pm$0.08 &0.33$\pm$0.06 &0.29$\pm$0.05 &0.45$\pm$0.11 &0.43$\pm$0.04 &375/288\\
\end{tabular}
\end{center}
$^a$  the radius from the center of NGC 1399\\
 $^b$  the ratio of emission measure of the lower to higher temperature components\\
$^c$  the single-temperature vAPEC for the ICM\\
$^d$  the two-temperature vAPEC model for the ICM\\
\end{table*}

\begin{figure}
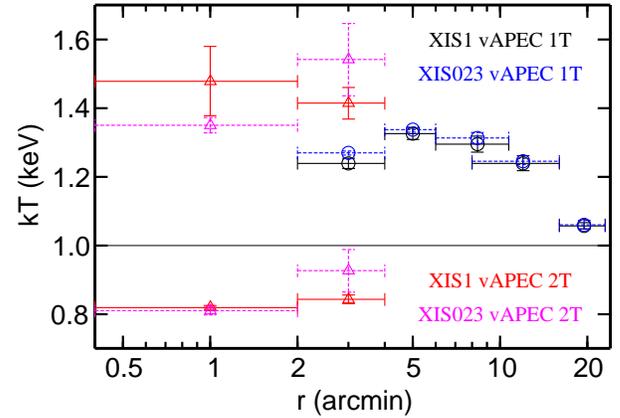

  \begin{center}
    \FigureFile(80mm,40mm){PASJ2938_figure4.ps}
  \end{center}
  \caption{The temperature profile of the ICM using the single
 temperature vAPEC model (open circles) and the two temperature vAPEC model (open triangles)
derived from XIS1 (solid lines) and XIS0,2,3 (dotted lines).}\label{fig:kT}
\end{figure}

The temperature of the ICM is almost constant at 1.3 keV from a radius
of $4'$ to $16'$ and outside $16'$, it drops to 1 keV (Figure \ref{fig:kT}).
Within $4'$, we need two-temperature components, 0.8 keV and 1.4 keV, to
fit the spectra.

\begin{figure}
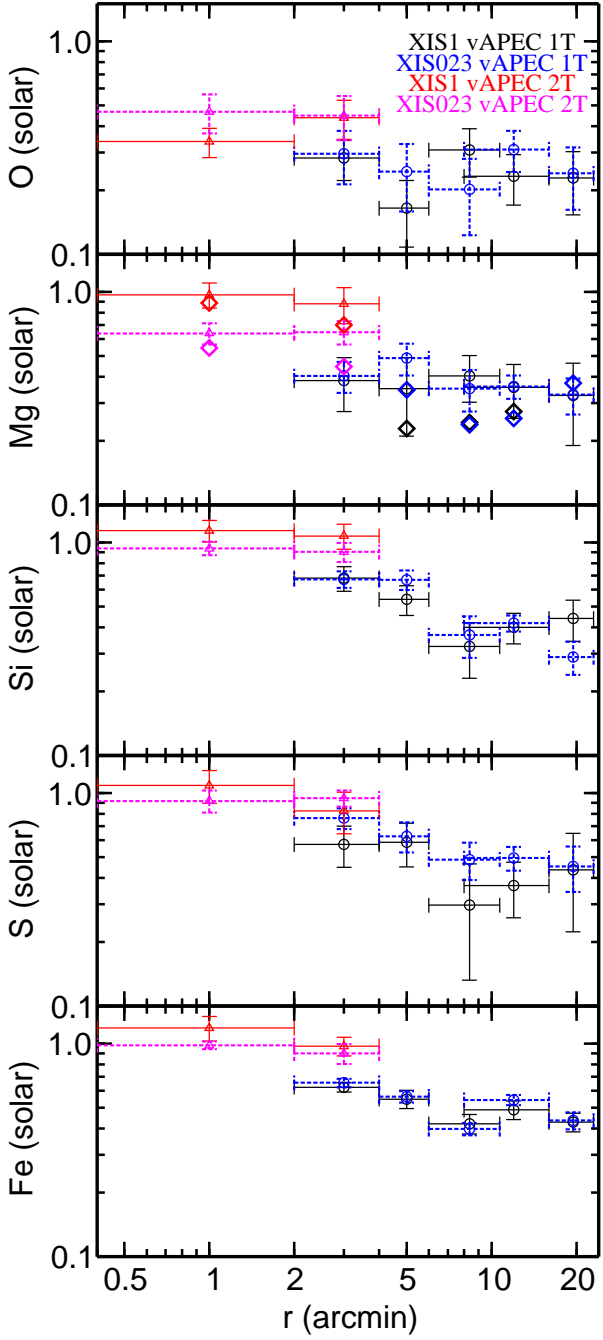

  \begin{center}
  \FigureFile(80mm,40mm){PASJ2938_figure5.ps}
\end{center}
\caption{The abundance profiles of O, Mg, Si, S and Fe based on the
single temperature vAPEC model (open circles) and the two temperature
 vAPEC model
 (open triangles)
derived from the XIS1 (solid lines) and XIS0,2,3 (dotted lines)
detectors.   In the Mg
abundance panel, the diamonds
 are the best-fit values derived from the vMEKAL model (see text).
}\label{fig:Fe}
\end{figure}
\begin{figure}
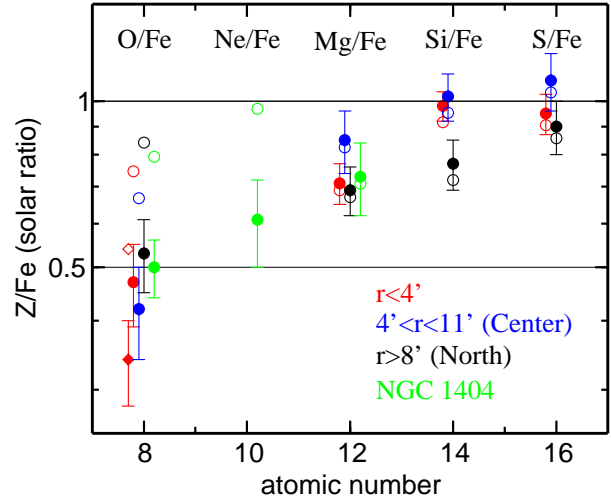

  \begin{center}
 \FigureFile(80mm,40mm){PASJ2938_figure6.ps}
\end{center}
    \caption{Abundance ratios of various elemental species compared to Fe
    for 3 radial regions: $r<4'$ (red), $4'<r<11'$ for the central
    field (blue), $r>8'$ in the north field (black), and NGC 1404
    (green).  Only the $r<4'$ data are fitted with a two
    temperature model.   For the O/Fe ratio, the closed circle and
 closed diamond correspond to $r<2'$ and $r>2'$, respectively.
 The open symbols are the abundance ratios using the new
solar abundance by Lodders (2003).
The error bars do not include systematic
    errors, which are estimated to be $\sim 20\%$ for the O/Fe and
    Mg/Fe ratios.  }
\label{fig:Z_vs_ZFe}
\end{figure}

Figure \ref{fig:Fe} shows the radial abundance profiles of O, Mg, Si,
S and Fe.   Adopting the two-temperature model at $r<4'$, the Fe abundance
is solar in the center, then decreases to about 0.5 solar and stays at
this level beyond a radius of $6'$. 
 The single temperature model gives significantly higher values of $\chi^2$
and a factor of 1.5--2 lower values of abundances, while the abundance
ratios are consistent with those from the two temperature fits.
 Si and S show almost the same
values as Fe, and these abundance values are consistent with those
obtained with XMM-Newton (Buote et al.\ 2002).
O, Mg, Si, S abundances at $r<4'$ agree within 10\% with those
derived from Chandra data (Humphery and Buote 2006).
  On the other hand, the Mg
abundance is 20-30\% lower and O is a factor of 2--3 lower compared
with the level of Si, S, and Fe. 
 Adding  temperature components  did not change
the abundance ratios although the error bars of the absolute abundances
became larger.
 Figure \ref{fig:Z_vs_ZFe} summarizes
the abundance ratios of O, Mg, Si, and S divided by the Fe value in
units of the solar ratio.  The O
abundance within a radius of $2'$  may have a large systematic uncertainty, since the spectral fit
around the O line had a problem.  The abundance ratios are consistent
with having no radial gradients.  On the average, O/Fe, Mg/Fe, Si/Fe, S/Fe
values are 0.47$\pm$0.05 (excluding r$<$2'), 0.72$\pm$0.05, 0.93$\pm$0.05, 0.96$\pm0.05$ in solar units, clearly 
indicating that the ratio grows as a function of atomic number.

\subsection{Uncertainties in the spectral fits}

In this section, we study  uncertainties in the abundance determination due
to uncertainty in the spectral models and
uncertainty in the calibration of the detectors.

\begin{figure}
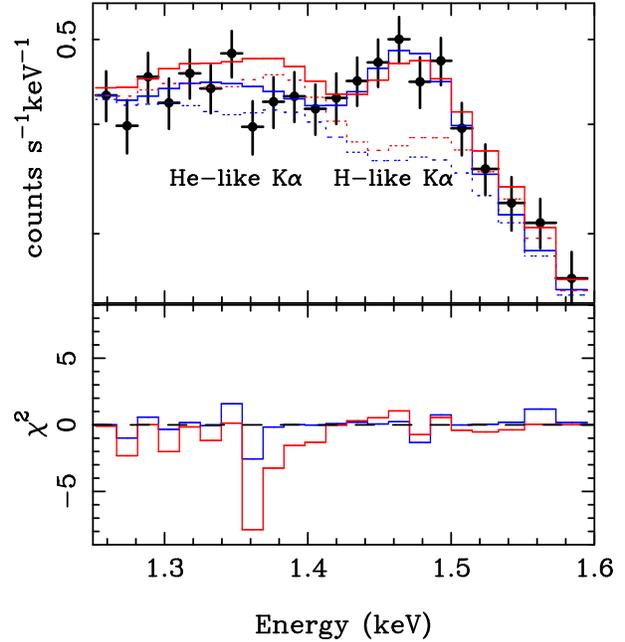

  \begin{center}
 \FigureFile(80mm,40mm){PASJ2938_figure7.ps}
  \end{center}
  \caption{The XIS0+2+3 spectrum for $4'<r<6'$ near the Mg K-shell lines
  (sold circles) fitted with the vAPEC (blue) and vMEKAL models
  (red).  The best-fit Mg abundances are 0.49 solar and 0.29 solar
  for the vAPEC and vMEKAL models, respectively. The dotted lines show the
  same best-fit models, with the Mg abundance set to zero.
  The lower panel shows the contribution to the total $\chi^2$ from each energy bin}
\label{fig:Mgspec}
\end{figure}

In order to look into the systematic effect by the plasma code on the
abundance determination, we fitted the spectra with the vMEKAL model (Mewe et al. 1995, 1996; Kaastra
1992; Liedahl et al. 1994) and
compared the results with those from the APEC model.  The reduced
$\chi^2$ from the vMEKAL model are systematically larger than the
vAPEC case, since the latter model gives improved fit for the Fe-L
lines.  As a result, the Mg abundance from the vMEKAL model is 20--40\%
lower than the vAPEC fit, although the temperature and abundances of
O, Si, S, and Fe agree within 10\%.  Figure \ref{fig:Mgspec} shows the
representative spectrum around the Mg K-shell lines.  The best-fit vAPEC model
and the observed spectrum agree well in this energy range, while the
vMEKAL model tends to over predict  the He-like Mg line.
  The $\chi^2$
values for the limited energy range 1.25--1.45 keV is 7.3 and 18.7 for 14
bins for the vAPEC model and vMEKAL model, respectively.
 Therefore, the Mg abundance
derived from the vAPEC model is more reliable than the vMEKAL fit, and
the systematic uncertainty in the Mg abundance concerning the atomic
data is thought to be much less than 40\%.

The true value of the metal abundance at the center may be higher,
because of projection effects and the point spread function of the
X-ray telescope, both of which tend to dilute the central sharp feature.  The
XMM-Newton data, fitted with a two temperature model, show the Fe
abundance within 20 kpc to be 1.5--2 solar (Buote et al.\ 2002).  We
fitted the central $r<2'$ spectrum with 2 APEC models with the
temperature and abundance of one component fixed at the best-fit
values for $4'<r<6'$.  The resultant abundance ratios remained the
same, but higher abundances by several tens of percent were allowed.

\begin{figure}
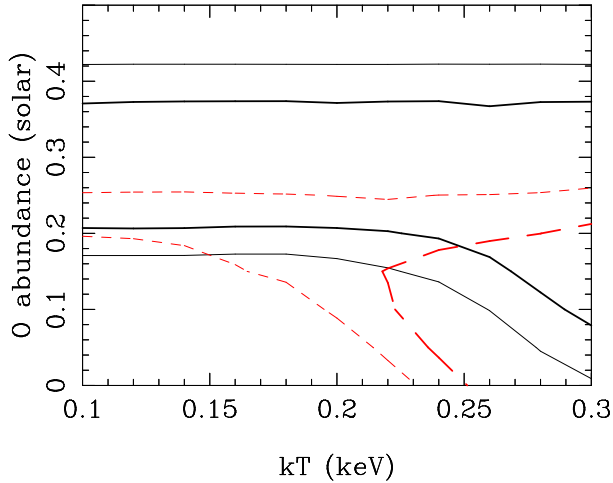

  \begin{center}
 \FigureFile(80mm,40mm){PASJ2938_figure8.ps}
  \end{center}
  \caption{Confidence contours (90\%, and 99\%) for the O abundance against the temperature of
 the MWH for the spectral data at $8'<r<16'$ (black solid lines) and
 $r>16'$ (red dashed lines). }\label{fig:cont}
\end{figure}

Since the Milky Way emits in O, modeling uncertainties could cause 
systematic uncertainty in  determining the O abundance of the ICM.
The MWH component has a higher temperature than the LHB and therefore,
may contribute to the H-like O line.
To investigate this effect, we added  an APEC model for the MWH
and repeated the spectral fit.
Figure \ref{fig:cont} shows
the O abundance versus the assumed temperature of the MWH.  
For the data in the outermost region ($r>16'$) 
 at the average temperature, 0.20 keV, of the MWH (Lumb
et al.~2002), the O abundance has not changed very 
much from the results in \S3.1. However,
when the temperature of the MWH is higher
lower O abundances are allowed, while the
upper limit of the O abundance does not change.
For the data in the $16'>r>8'$, even when  the 
temperature of the MWH is 0.20 keV, 
lower O abundances are required.

The chemical composition, column density and uniformity of the contaminant
on the XIS filter are still poorly known, and therefore, we have
to carefully check its effect on the abundance determination.  Outside
a radius of $2'$, the column densities of C derived from our
spectral fit mostly agree  with those determined from the hardware
calibration  within a systematic error of several times
10$^{17}{\rm cm^{-2}}$ (Koyama et al.\ 2006).  However, within $2'$, the spectral
fit gives significantly higher column densities for the FI detectors
 by $1\sim2\times 10^{18}{\rm cm^{-2}}$. 
This discrepancy may be related to the astrophysical spectral model,
 since all of the FI detectors give a similarly high value of
the column density.  When we fit the spectra by fixing the column
density to the value given by the contamination model from the hardware 
calibration, with the hydrogen
column density allowed to vary, the O abundance increases by 10\%.

The contaminant is not spatially uniform on the filter.  In the north
field, the regions where we accumulated the spectra have 
different column densities.  To check the effect, we divided the
region of $r>8'$ in the north field into two regions: namely, outside
and inside of the radius of $6'$ from the detector center. We derived
similar O and Fe abundances for the two
regions, verifying that the spatial non-uniformity of the contaminant is not a
significant problem.

Since the O/C ratio of the contaminant can be smaller than our current best 
estimate of 1/6, we also fitted the spectra using ratios of
1/10 and 1/20. This gave systematically lower O abundance than the
nominal case by 10\% and 20\% for the O/C ratio of 1/10 and 1/20,
respectively. The temperature and abundances of the other elements did
not change significantly.

In summary, uncertainty in Fe-L modeling may cause 20--30\%
systematic errors in the Mg abundance.  If the temperature of the
MWH is higher than $\sim $0.2 keV
the O abundance becomes lower in the outer regions. The
uncertainty concerning the thickness and chemical composition of the
XIS contaminant results in $\sim 20\%$ systematic error in the O
abundance.
However, the observed constancy with radius of the
derived ratio of Mg to O, which are both  mostly synthesized in SN II
indicates that our O and Mg abundance may  not be strongly affected by 
these uncertainties.

\subsection{NGC 1404}

We accumulated an on-source spectrum of NGC 1404 within a radius of $3'$
centered on the galaxy.  The background spectrum was taken from the
Lockman-Hole observation over the same detector region.
 Since the surface brightness of NGC 1404 is high, uncertainty in the 
Galactic emission is negligible.
In order to subtract the local ICM component, we took a spectrum just
outside of the on-source region in a radius $3'-4.2'$, from which we
subtracted the Lockman-Hole spectrum integrated over the same detector
coordinate.  The spectrum for this ring region was fitted with a model
consisting of 2 vAPEC components: one for the ICM and the other for
the ISM of NGC 1404 with a fixed temperature of 0.6 keV\@,
 which was derived from the spectral fit of the on-source region.
The temperature of the ICM component was
derived to be $1.44\pm 0.10$ keV\@.  Using this temperature, we
returned to the on-source ($r< 3'$) spectrum and fitted it with a 3
component model: 2 vAPEC models and a thermal bremsstrahlung. Two
thermal components are for ISM and ICM, with the ICM temperature and
abundance fixed at the values of the previous ring-region fit. The
thermal bremsstrahlung with a fixed temperature of 10 keV represents
discrete source contribution.  As for the column density of the
contaminant, we assumed the expected value from the calibration
(Koyama et al.\ 2006) and the hydrogen column density was allowed to
vary freely.  The spectra for the XIS0, XIS2, and XIS3 detectors were
fitted simultaneously.

\begin{table*}[t]
\caption{Spectral fit results for NGC 1404.}\label{n1404}
\begin{center}
  \begin{tabular}[t]{lcclllllll}
 model & XIS  & kT & N$_{\rm{H}}$& O & Ne & Mg & Si & Fe &$\chi^2$/dof \\
      &           & (keV) & ($10^{20}/\rm{cm^{2}}$)& (solar) & (solar) &
   (solar) & (solar) & (solar) &\\
vAPEC 1T$^a$ & 023$^b$ & 
0.60$\pm$0.01 & 0.6$\pm$ 0.6 &0.44$\pm$0.08 &0.71$\pm$0.14 &0.78$\pm$0.13 &0.71$\pm$0.18 &1.07$\pm$0.20 &619/467\\
vAPEC 1T & 1 &  0.59$\pm$0.01 & 2.4$\pm$ 0.6 &0.42$\pm$0.05
   &0.46$\pm$0.11 &0.58$\pm$0.11 &0.96$\pm$0.14 &0.79$\pm$0.04
   &215/154\\
vAPEC 2T$^c$ & 023 & 0.60$\pm$0.01 & 2.1$\pm$ 2.1 &0.39$\pm$0.06
   &0.61$\pm$0.09 &0.66$\pm$0.11 &0.58$\pm$0.11 &0.96$\pm$0.15
   &618/465\\
vAPEC 2T        &  023 &  ---$^d$ & \\
vAPEC 2T & 1 & 0.59$\pm$0.01& 3.1$\pm$ 1.1 &0.37$\pm$0.08
   &0.76$\pm$0.16 &0.89$\pm$0.15 &1.26$\pm$0.24 &1.18$\pm$0.09
   &204/151\\
vAPEC 2T  & 1 &  0.27$\pm$0.02\\
vMEKAL 1T & 023 &  0.59$\pm$0.01 & 3.9$\pm$ 1.0 &0.31$\pm$0.04 &0.35$\pm$0.06 &0.41$\pm$0.05 &0.35$\pm$0.06 &0.52$\pm$0.02 &732/461\\
vMEKAL 1T & 1 &  0.58$\pm$0.01 & 2.6$\pm$ 0.5 &0.39$\pm$0.05
   &0.39$\pm$0.11 &0.66$\pm$0.10 &0.87$\pm$0.15 &0.71$\pm$0.06
   &229/154\\
 \end{tabular}
\end{center}
$^a$ The single temperature model for the ISM\\
$^b$ The simultaneous fit of the spectra of XIS0, XIS2, and XIS3\\
$^c$ The two-temperature model for the ISM\\
$^d$ not constrained\\
\end{table*}

\begin{figure}
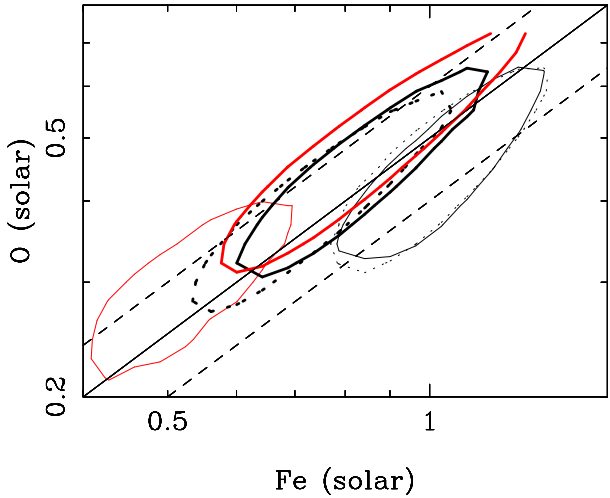

  \begin{center}
    \FigureFile(80mm,40mm){PASJ2938_figure9.ps} 
 \end{center} 
    \caption{The 90\% confidence contours of O abundance against
    Fe abundance for NGC 1404. The thin curves are for XIS0,2,3,
    and the thick ones for XIS1. Solid black curves show the single temperature
    vAPEC fit, and solid red  curves are for the vMEKAL case. The black dotted
    contours were derived while allowing the contaminant thickness  to
    vary.  The solid line shows an O/Fe ratio 0.5 and the dashed
    lines denotes ratio values of 0.4 and 0.6.  }
\label{fig:n1404ovsfe}
\end{figure}

The results are summarized in Table \ref{n1404}. 
The temperature of the ISM is determined to be 0.6 keV\@.  Except for
the Si abundance, the results from the BI and the FI detectors are
mostly consistent.  The abundances of O, Ne, Mg, Fe become higher in
order, ranging from 0.5 to 0.9 solar. 
Figure \ref{fig:n1404ovsfe} shows the confidence contours of the O
abundance against the Fe abundance.  The elliptical shape of the
contour indicates that the abundance ratio is better determined than
the abundances themselves.
 The abundance ratios among
O, Mg and Fe are close to those obtained for NGC 1399 and the ICM
(Figure \ref{fig:Z_vs_ZFe}).  The Si abundance has a large systematic
uncertainty, since the dominant He-like Si line at $kT = 0.6$ keV
falls in the energy where the response matrix has a problem.

\begin{figure}
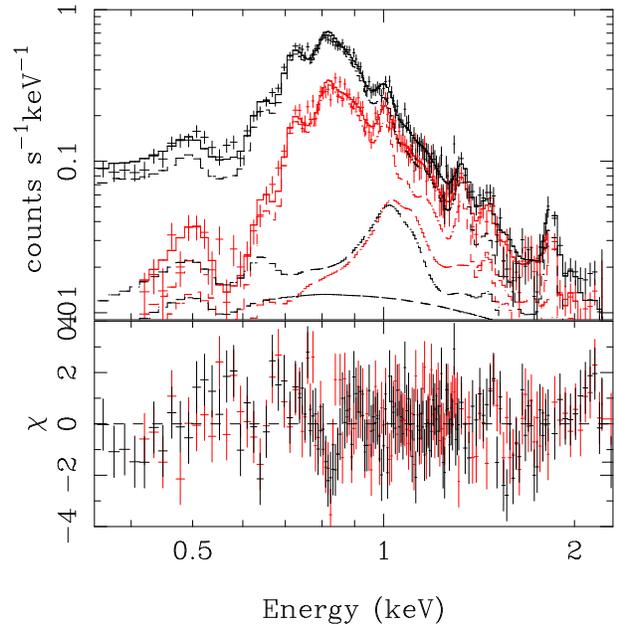

  \begin{center}
    \FigureFile(80mm,40mm){PASJ2938_figure10.ps}
  \end{center}
  \caption{XIS1 (black) and the XIS0 (red) spectra of NGC 1404
 fitted with a 3-component model: 2 vAPEC models for the ISM and ICM, and a
 bremsstrahlung for discrete sources. The contribution of each
 component is shown with a dashed line.  }\label{fig:n1404spec}
\end{figure}

The spectra of
XIS0 and XIS1 are shown in Figure \ref{fig:n1404spec}.  The reduced
$\chi^2$ values are 1.3--1.4, and there are residual structures
around 0.7--0.8 keV which are likely to be related to poorly modeled Fe-L lines.
The residual structures are seen in the fits for all detectors, but are
not easily recognized in the spectra of NGC 1399 and ICM where the
temperature is higher. 
These discrepancies in the Fe-L energy range are also seen in the RGS
spectrum of 
the X-ray luminous elliptical galaxy, NGC
4636, whose ISM temperature is  also $\sim$ 0.6 keV
 (Xu et al.\ 2002) .
A two-temperature vAPEC model for the ISM did not improve the
reduced $\chi^2$ for the NGC 1404 fit very much. 
 Therefore, there might still be some problem in the Fe-L atomic data.

The reduced $\chi^2$ value for a single temperature MEKAL model for the ISM is
higher than the vAPEC fit and the derived Fe abundance is  lower.
However, abundance ratios of O, Ne, Mg and Fe agree within 10--20\% between the two
models.  Therefore, systematic uncertainties in the O, Ne and Mg
abundances due to the uncertainty in the Fe-L atomic data may not be large.

We also tried the fit with different background, which consists of
data for dark earth, a ring-like region with $r =3'-4.2'$ just outside
the on-source region, and a region near the edge of the detector.
Then, the best-fit values of the abundances have changed by several
tens of percent, while the abundance ratios
remained the same.

\begin{figure}
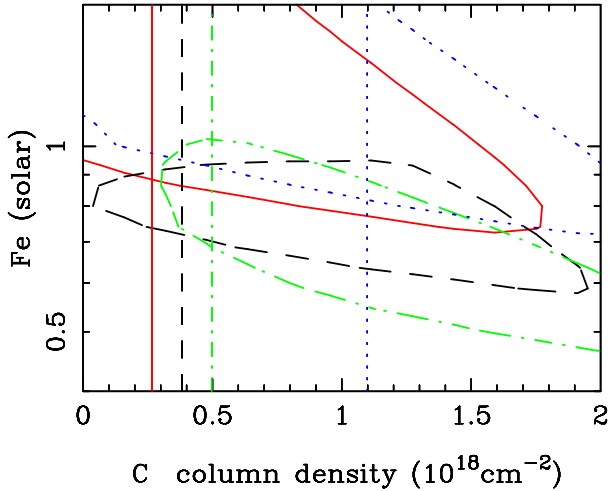

  \begin{center}
    \FigureFile(80mm,40mm){PASJ2938_figure11.ps}
  \end{center}
  \caption{The 90\% confidence contours of Fe abundance against
   column density of C in the contamination. Colors denote XIS0
  (red solid), XIS1 (black dashed), XIS2 (green dot-dashed), and XIS3
 (blue dotted).
  The vertical lines correspond to the standard thickness
 of the contaminant from Koyama et al.\ (2006)}\label{fig:cvsfe}
\end{figure}

At the temperature of 0.6 keV, the abundances are mainly determined by
the ratio of line strengths to the continuum level below 0.6 keV\@.
We allowed the contaminant column density to vary and fitted the
spectrum of each detector with the same model. The Fe abundance decreased
as we increased the column density (Figure \ref{fig:cvsfe}). 
As shown earlier, adopting the standard thickness of the contaminant from
Koyama et al.\ (2006), the Fe abundance is about solar for all 
detectors.

The metal abundance of the hot ISM of NGC 1404 was first derived with
ASCA (Loewenstein et al.\ 1994), and an extremely low value
of $\sim$0.2 solar was obtained.  With ASCA, we were unable to determine the continuum
level below 0.6 keV\@. Also, there was a problem in the atomic data of
Fe-L lines, and an incorrect assumption about the abundance ratio resulted
in the very low abundance value (Arimoto et al.\ 1997, Matsushita et
al.\ 2000).  With Suzaku, we can detect both the continuum below 0.6
keV and the peak of the O line, and, therefore, the systematic
uncertainty has been much reduced.

In summary, the Fe abundance of the X-ray emitting gas of NGC 1404 is
shown to be about solar, and the O/Fe, Ne/Fe, and Mg/Fe ratios are
0.50$\pm$0.06, 0.61$\pm$0.11, and 0.73$\pm 0.11$, respectively, in
solar units. 
 The uncertainty in the thickness of the contaminant and the
background give systematic error in the Fe abundance by several tens
of percent, while they do not affect the abundance ratios.

\section{Discussion}

\subsection{Summary of the abundance determination}

Suzaku observations of the Fornax cluster have revealed  abundance profiles
of O, Mg, Si, S and Fe in the ICM for a wide region extending 130 kpc
north and 60kpc south of the cD galaxy, NGC 1399.  Also, in NGC
1404, abundances of O, Ne, Mg and Fe in the ISM were derived.
Hereafter, we call the region within a radius of $4'$ from NGC 1399 as the
NGC 1399 region, and outside of this circle as the ICM region.  The Fe
abundance in the NGC 1399 region and in NGC 1404 is about solar, and
it decreases to 0.5 solar in the ICM region.  The abundance ratios
among O, Mg, Si, S and Fe show common values for all 3 regions
(Figure \ref{fig:Z_vs_ZFe}). 
 On average, O/Fe, Mg/Fe, Si/Fe, and S/Fe
ratios of the NGC 1399 region and the ICM region are 
0.47$\pm$0.05 (excluding r$<$2' where the spectral fittings have
problems around O lines), 0.72$\pm$0.05, 0.93$\pm$0.05,
0.96$\pm0.05$ in solar units,
respectively.  O/Fe, Mg/Fe and Ne/Fe ratios for the ISM of NGC 1404 are
0.50$\pm$0.06, 0.61$\pm 0.11$ and 0.73$\pm0.11$, respectively. 
 Adopting new solar abundances by Lodders (2003),  O/Fe ratios
of the NGC 1399 region,  the ICM region and NGC 1404  increases to
0.81$\pm$0.09 and 0.87$\pm $0.1 in solar units, respectively.
 Ne/Fe ratio in NGC 1404 also increases to 1.0$\pm 0.2$ in solar units
(Figure \ref{fig:Z_vs_ZFe}).
These
abundance ratios are similar to those observed around cD galaxies and
in elliptical galaxies (e.g., Matsushita et al.\ 2003; 2006; Tamura et
al.\ 2003; Xu et al.\ 2002).
 However,  the observed constant O/Fe and Mg/Fe ratios differ from other
systems in which a radial increase in the O/Fe ratio has possibly been seen.
For example, the O/Fe ratio of the ICM increases by several tens of 
percent within 80 kpc
for M 87 (Matsushita et al.\ 2003).  The other systems have poorer
statistics, but the O abundance spatial variation seems flatter than 
Fe (e.g., Tamura et al.\ 2004).

 In the remainder of this section, we discuss 
the abundance pattern of O/Ne/Mg and O/Si/Fe and consider nucleosynthesis by 
SN Ia and SN II (section 4.2 and 4.3).
In section 4.4, the O and Mg abundances of the NGC 1399
region and NGC 1404 are compared to
the stellar metallicity of these galaxies.  Section 4.5 considers 
the Fe abundance of these galaxies in light of their current estimated
SN Ia rates.
Finally, the O and Fe mass-to-light ratio in the ICM region are derived
in section 4.6, where we also discuss the origin of the metals in the ICM.

\subsection{Abundance pattern of O/Ne/Mg and nucleosynthesis of SN II}

\begin{figure}
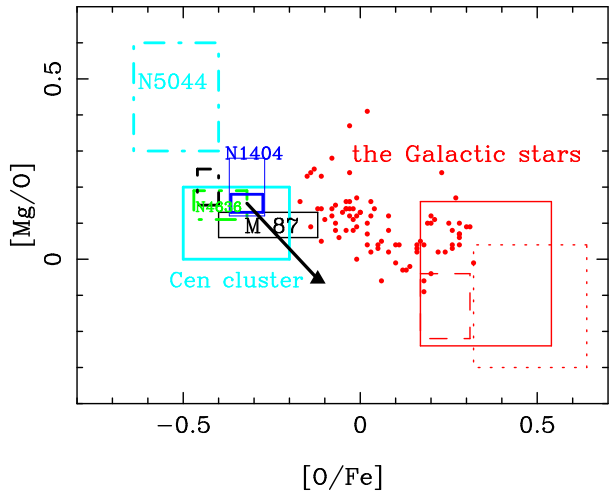

  \begin{center}
    \FigureFile(80mm,40mm){PASJ2938_figure12.ps}
  \end{center}
  \caption{[Mg/O] of the NGC 1399 region and the ICM region (thick blue
 lines), NGC 1404 (thin blue lines), NGC 4636 (green; Xu et al. 2002), M
 87 (black; Matsushita et  al. 2003;  Werner et al. 2006),
the Centaurus cluster (light blue; Matsushita et al. 2006), and NGC 5044
 (light blue Tamura et  al. 2003) are plotted against [Fe/O].
The RGS and EMOS results are plotted in dot-dashed and solid lines, respectively.
 The arow denotes abundance correction,  adopting 
 the new solar abundance model by Lodders (2003). 
Those of disk stars by Edvardsson et al. (1993; red closed circles) 
 and
Clementini et al. (1999; red dotted box)
 in the Galaxy   are also plotted.
 The red solid and dashed boxes correspond to mean values of thick and thin disk stars
in the bin of [Fe/H] -0.45 to -0.55 by Reddy et al. (2006)
}\label{fig:mgo}
\end{figure}

Since SN Ia are not significant sources of O, Ne, and Mg, the abundance
pattern of these elements can be used to infer the  contribution from 
SN II\@.  The
observed Mg/O ratio is consistent among NGC 1404, the NGC 1399 region,
and the ICM region, and the average value is 1.5 in  solar units.
Considering the systematic uncertainty of 20--30\%, this value is
close to the level in other systems,   M 87 (Matsushita et al. 2003;
Werner et al. 2006),
the Centaurus cluster (Matsushita et al. 2006), NGC 4636 (Xu et
al. 2002), and NGC 5044 (Tamura et al. 2003), 
 as summarized in Figure \ref{fig:mgo}, although
  the Mg/O ratio for NGC 5044 is
larger than 2.
We selected these groups and clusters observed with XMM-Newton
for comparison, since  they are  all giant elliptical galaxies
and the Mg/O ratio of the gas surrounding them reflects the
composition of their stars. In addition, they are among the
  brightest objects with well observed spectra clearly showing  
 O emission lines from the EMOS (M 87 and the Centaurus cluster) 
and RGS ( M 87, NGC 4636 and NGC  5044).
 These Mg/O
ratios for the cD and elliptical galaxies reflect the metallicity of
the stars. Also, the result suggests that possible old SN II ejecta
from early-type galaxies may be the main component of these elements
in the ICM region.

 In order to compare nucleosynthesis in elliptical galaxies and spiral
galaxies, it is worth to compare the abundance pattern in the ICM and
ISM of elliptical galaxies with that of stars and ISM in our Galaxy.
It is important to bear in mind that the methods of abundance determination 
for these various classes are very different
which can introduce some systematic uncertainties in the comparison.
 Adopting the new solar abundance by Lodders (2003), the
X-ray--derived Mg/O ratio decreases by a factor of 1.6
and are  consistent  with those of Galactic disk stars
(Figure \ref{fig:mgo}; Edvardsson et al.\ 1993; Clementini et al. 1999; Reddy et al. 2006).  
The Ne/O and Mg/O ratios in the ISM of the
Galaxy are measured by X-ray absorption, and the obtained values are
1--2 (Yao \& Wang 2006) and 1.3--2.3 (Ueda et al.\ 2006),
respectively.  Therefore, at least concerning the ratios of O/Ne/Mg,
there is no obvious difference in the nucleosynthesis products of SN
II between elliptical galaxies and the Galaxy.

\subsection{Abundance pattern of O/Si/Fe}

\begin{figure}
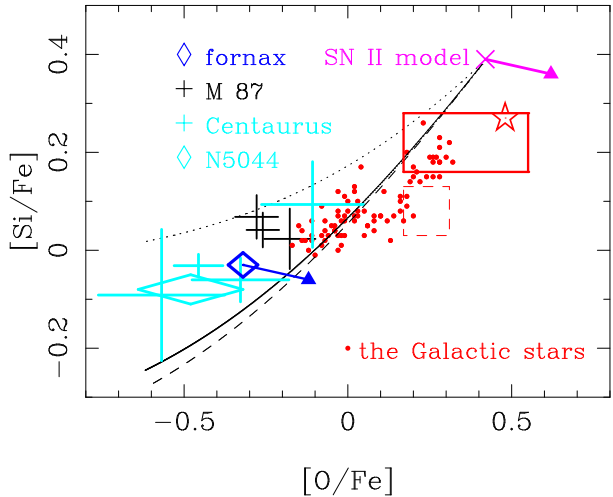

  \begin{center}
    \FigureFile(80mm,40mm){PASJ2938_figure13.ps}
  \end{center}
  \caption{[Si/Fe] of the NGC 1399 region and the ICM region (thick
 blue diamond),   M 87 (black crosses;Matsushita et al. 2003),
the Centaurus cluster (light blue crosses;Matsushita et al. 2006), and NGC
 5044 (light blue diamonds; Buote et al. 2003) are plotted against [O/Fe].
 The arows denote abundance corrections,  adopting 
 the new solar abundance model by Lodders (2003). 
Those of disk stars (Edvardsson et al. 1993; closed circles), 
and average of metal poor  stars (Clementini et al.\ 1999, star)
and in the Galaxy   are also plotted. 
 The solid and dashed boxes correspond to mean values of thick and thin disk stars
in the bin of [Fe/H] -0.45 to -0.55 by Reddy et al. (2006).
The SN II nucleosynthesis model described in Iwamoto et al.\ (1999) is
 show in the cross.
The solid line, dotted line, and dashed line represent a sum of  SN II nucleosynthesis model and W7
 model, WDD1 model, and  WDD3 model, respectevely. 
}\label{fig:osife}
\end{figure}

The abundance ratios for O/Si/Fe in the NGC 1399 region and 
 the ICM region and for O/Fe
in NGC 1404 are close to the previous measurements around  brightest cD 
galaxies
and in the central galaxies of groups of galaxies, including M 87
(Matsushita et al.\ 2003), the Centaurus cluster (Matsushita et al.\
2006), NGC 4636 (Xu et al.\ 2002), and NGC 5044 (Tamura et al. 2003),
 as summarized in Figure \ref{fig:osife}.
Si abundances of the Fornax cluster is 0.1 dex 
higher than   a sum of SN II nucleosynthesis model described in Iwamoto et al. (1999) and 
 classical deflagration SN Ia model,  W7 (Nomoto et al. 1984), 
which   predicts  an Fe/Si abundance ratio of 2.6.
Those of other clusters are also 0.1--0.2 dex higher than the sum.

 In order to account for the high Si abundance, Finoguenov et al.\
 (2002) and
 Matsushita et al. (2003) suggested that  ejecta of SN
Ia should have an Fe/Si ratio of about 1 solar.
Matsushita et al. (2006), Buote et al. (2003), Hampherey and Buote
 (2006) also discussed higher Si abundance production by SN Ia, 
although some Chandra data in Hampherey and Buote (2006) 
are consistent with W7 model.
This Fe/Si ratio may be explained by
the explosion models.  A higher fraction of Si is indicated in the delayed
detonation  model, WDDs ( Iwamoto et al.\ 1999), which 
 give a range of Fe/Si ratios from 1 to 3, which might
be related to the age of the system (Umeda et al.\ 1999). 
 A sum of the   SN II nucleosynthesis model plus the WDD1 model is consistent 
with data with the highest Si/Fe ratios, while nearly all of the data points
lie between this model and the low Si producing SN Ia nucleosynthesis models 
(e.g, W7 or WDD3).

Another reference of the  total nucleosynthesis from SN II in our Galaxy
is  the abundance pattern of metal poor Galactic stars.
 Since the SN II nucleosynthesis model might have some systematic
uncertainty, it is important to compare the abundace pattern of the
ICM and a sum of  the metal poor Galactic stars and SN Ia model,
although  abundances of stars might have 0.1--0.2 dex uncertainty
as described in Asplund (2005).
Adopting the new solar abundance model in Lodders (2003),
the [O/Fe] of ICM  increases by 0.2 dex (Figure \ref{fig:osife}).
Then, the  O/Si/Fe pattern of the ICM becomes more consistent with a sum of 
the Galactic metal poor stars and W7.

\subsection{Stellar metallicity in elliptical galaxies}

In elliptical galaxies, O and Mg abundances in the hot gas should be
equal to those in mass losing stars, since these elements are not
synthesized to any great extent by SN Ia.  Therefore, X-ray observations 
can probe the
stellar metallicity over the entire galaxy, which is virtually impossible
with optical observations.  
 The stellar metallicity of elliptical galaxies is usually studied
by optical spectroscopy using the Mg absorption line index (e.g., Kobayashi \& Arimoto 1999),
which depends not only on the  Mg abundance but also on the total
metallicity where the O contribution matters the most.
The problems are that the index also depends on the age distribution of
the stars, and  optical spectroscopy tends to be limited to within the 
central region of galaxies.
 In addition,  these values are based on abundance studies in the
Galactic stars, which might change when considering the uncertainties in
calculation of stellar atomosphere (Asplund 2005).

 Adopting the new solar abundance by Lodders (2003),
the O and Mg abundances of the ISM of NGC 1404 are about  0.7 solar.
Those of NGC 4636 are  0.8 and 0.65 solar (Xu et al. 2002), and therefore
the two galaxies should have similar stellar metallicity and in addition,
we would expect the O and Mg abundances of the stars in the two galaxies to
be 0.6--0.8 solar.
The Mg$_2$ index of NGC 1404 has only been measured in the galaxy's
center; Faber et al.\ (1989) quote a value of 0.317.
This is close to the central value of 0.311 seen in NGC 4636 whose
Mg$_2$ index has been observed out to $\sim 1 r_e$.
Considering the gradient of the index, the extrapolated stellar
metallicity of NGC 4636 is about 0.7 solar (Kobayashi \& Arimoto 1999).
Thus in NGC 4636 the ISM metallicity and stellar metallicity 
derived from the Mg$_2$ index are consistent.
Assuming that the stars in NGC 1404 have a similar metallicity
gradient to NGC 4636, we can also conclude that the ISM metallicity
and stellar metallicity should be consistent as well.

The central Mg$_2$ index of NGC 1399 is 0.344, which is higher than
those of NGC 4636 and NGC 1404.
The O and Mg abundance of the NGC 1399 region are consistent or higher
than those of NGC 4636 and NGC 1404, since the absolute value of abundance of the NGC 1399 may be higher 
by several tens of percent when one considers projection.

\subsection{SN Ia nucleosynthesis in elliptical galaxies}

 The  observed  Fe abundance in the ISM of NGC 1399 and NGC 1404
 can constrain the present metal supply from galaxies to the
ICM\@. 
 Assuming the SN II abundance pattern by Iwamoto et al. (1999),
 $\sim$ 80\%of Fe  and $\sim$ 40\% of Si are synthesized by SN Ia.
 The Fe abundance enriched by SN Ia in an elliptical galaxy is
proportional to $M^{\rm Fe}_{\rm SN} \theta_{\rm SN}/\alpha_*$ (see
Matsushita et al.\ 2003 for details).  Here, $M^{\rm Fe}_{\rm SN}$ is
the Fe mass synthesized in one SN Ia, $\theta_{\rm SN}$ is the SN Ia
rate, and $\alpha_*$ is the stellar mass loss rate, respectively.  Fe
in the hot ISM of these galaxies is mainly produced by SN Ia, since SN
II synthesize much more O and Mg than Fe.  
  We used the mass loss rate from Ciotti et al.\ (1991) assuming the age to
 be 13 Gyr,
which is  approximated by $1.5\times10^{-11}L_B
t_{15}^{-1.3} M_\odot /yr$, where $t_{15}$ is the age in unit of 15 Gyr,
and $L_B$ is the B-band luminosity. 
 $M_{\rm Fe}$ produced by one SN Ia explosion is likely to be $\sim
 0.6M_\odot$ (Iwamoto et al. 1999). 
Combining with the SN Ia rate optically observed by
Capperallo et al.\ (1997) which is 
0.13$\pm$0.05 ${h_{75}}^2$ SN Ia/100yr/10$^{10}L_B$, 
the resultant Fe abundance only considering
the SN Ia contribution is  2--4 solar.   This value is higher than
the presently observed Fe abundance, indicating a lower SN Ia rate.

\subsection{Mass to light ratio for O and Fe in ICM and origin of  the metals}

Since metals in the ICM are all synthesized in galaxies, the
metal-mass-to-light ratios are important to study the chemical
evolution of the ICM\@.  Within a radius of 130 kpc, or $0.13
r_{180}$, most of the optical light comes from NGC 1399 and several
other bright galaxies, including NGC 1404.  The total luminosity of
NGC 1399 has an uncertainty since the surface brightness of the cD
halo is very low.  Recent CCD photometry by Karick et al.\ (2003)
showed the B-band luminosity of NGC 1399 to be
$2.8\times10^{10}L_\odot$, and the total B-band stellar luminosity of
galaxies within 130 kpc to be $8\times 10^{10}L_{\odot}$.  Using 
old photographic data by Schombert (1986), Saglia et al.\ (2000)
derived the stellar luminosity profile of NGC 1399 out to nearly 1
degree from the galaxy center. At 130 kpc, the B-band luminosity of
NGC 1399 reaches $\sim 10^{11} L_{\odot}$.  Including the halo
component, the total B-band luminosity of all the galaxies within 130
kpc becomes $1.5\times 10^{11}L_{\odot}$.

The gas mass profile was derived by Paolillo et al.\ (2002) based on
the ROSAT data, and, within 130 kpc, the gas mass is estimated to be
$10^{11}\,M_{\odot}$.  As a result, the O mass-to-light ratio (OMLR) and Fe
mass-to-light ratio (IMLR) within 130 kpc are $2\times
10^{-3}\,M_{\odot}/L_{\odot}$ and $4\times 10^{-4}\,M_{\odot}/L_{\odot}$,
respectively.  These values are more than an order of magnitude lower
than those for richer clusters.  With ASCA, the IMLR for rich clusters was
measured to be $\sim$ 0.01$M_{\odot}/L_{\odot}$, while groups
of galaxies gave smaller values (Makishima et al.\ 2001),
 although angular resolution of ASCA often did not enough to 
obtain the  metallicity distribution and the obtained IMLR might be 
overestimated.
  With XMM, Tamura et
al.\ (2004) derived IMLR for several clusters within $\sim 250 h^{-1}$
kpc to be $\sim 0.01 M_{\odot}/L_{\odot}$.  Measurement of the OMLR for
rich clusters is not reliable (Tamura et al.\ 2004, Kawaharada 2006).
Our results can be compared with the Centaurus cluster,
whose temperature is 4 keV\@. Within a radius of $0.11 r_{\rm 180}$,
the OMLR and IMLR for the Centaurus cluster are   $3\times
10^{-2}M_{\odot}/L_{\odot}$ and $4\times 10^{-3}M_{\odot}/L_{\odot}$,
respectively, and
about an order of 
magnitude larger than those in the Fornax cluster (Matsushita et al.\
2006).

The total amount of Fe and O, when normalized by the stellar
luminosity and accumulated over the Hubble time, should be similar
among the clusters, unless much of these elements are lost from the
system. The constancy of the metal mass would be the case in the
Fornax cluster, as well as in rich clusters, since most of the stellar
light in the Fornax cluster comes from bright old galaxies just as in
rich clusters (Kuntschener 2000).  One difference between the poor and
rich clusters is that the gas in poor clusters and groups of galaxies
is more extended than in rich clusters compared with the stellar
distribution (e.g., Ponman et al.\ 1999).  Since the Fe abundance in
the ICM does not depend on the ICM temperature (Fukazawa et al.\
1998), the difference in the OMLR and IMLR should almost purely
reflect the difference in the gas fraction.
Since the hot gas in poor clusters and groups commonly show higher
entropy than the case of pure gravitational heating (e.g., Ponman et
al.\ 1999), the gas in poor clusters, such as Fornax, may have
expanded significantly. 
 The metal distribution in the ICM may be used
as a tracer of the history of such gas heating,
since both metal enrichment and heating timescales
determine the metal distribution in the ICM.

The O/Fe ratios in the Fornax cluster indicate that the Fe 
has been mostly produced by SN Ia.
If the gas and stellar distribution remain the same within 130 kpc
of NGC 1399, the accumulation time scale by SN Ia may be much
shorter than the Hubble time.  Therefore, significant amounts of O and
Mg has to come from stellar mass loss, although the time dependence of
the mass loss rate and SN Ia rate is not clear. 
 Therefore,   within a radius of $0.13r_{180}$ in the Fornax cluster,
the OMLR directly ejected from old SN II may be even lower.
 For example, in order to accumulate half of the observed OMLR with stellar
mass loss requires 7 Gyr, assuming that the O abundance of stars are 0.45
solar, the age of the
galaxies is 13 Gyr and using the
mass loss rate from Ciotti et al.\ (1991).
Therefore,  most of the O from old SN II may be lie beyond $0.13r_{180}$.

\section{Summary and conclusion}

With Suzaku observations of the Fornax cluster, the abundances of
O, Mg, Si, S and Fe of the ICM up to 130kpc north and up to 60kpc south
of NGC 1399, and O, Ne, Mg and Fe abundances of NGC 1404 were derived
accurately. 
The Fe abundances around NGC 1399 and NGC 1404 are about solar and
drop to 0.5 solar in the ICM.
The elemental abundance ratios show common values around NGC 1399, NGC 1404,
and the ICM: Si and S have similar abundances to Fe and
O/Fe, Ne/Fe, Mg/Fe ratios are 0.5--0.7 solar in units of the solar ratio.
The O, Ne and Mg abundances of NGC 1404 and the NGC 1399 region
are consistent with the stellar metallicity of these galaxies.
Most of the Fe around NGC 1399, NGC 1404 and the ICM should have been 
synthesized by SN Ia. The abundance ratios of O/Fe and Si/Fe in the gas 
are consistent with a mixture of SN Ia ejecta of the
 W7 model and metal-poor Galactic stars.
Values for the IMLR and OMLR are much smaller than in rich clusters,
which indicates that most of the metals may be outside of the region we 
observed.
The metal distribution of the ICM may be used as a tracer of the history
of the cluster.

\bigskip

 We would like to thank Prof. R. Kelley for careful internal
review of the manuscript.
The authors would like to thank ISAS/JAXA and the entire Suzaku
team for the opportunity to participate in the development of the
Suzaku mission as members of its Science Working Group.
This work is supprted by Grant-in-Aid for Science Research of JSPS.
JPH acknowledges support from NASA  grant NNG05GP87G.

\end{document}